\documentclass[pra,twocolumn,notitlepage,superscriptaddress,longbibliography]{revtex4-1}
\usepackage{amsmath}
\usepackage{amssymb}

\AtBeginDocument{%
    \newwrite\bibnotes
    \def\bibnotesext{Notes.bib}
    \immediate\openout\bibnotes=\jobname\bibnotesext
    \immediate\write\bibnotes{@CONTROL{REVTEX41Control}}
    \immediate\write\bibnotes{@CONTROL{%
    apsrev41Control,author="08",editor="1",pages="0",title="0",year="1"}}
     \if@filesw
     \immediate\write\@auxout{\string\citation{apsrev41Control}}%
    \fi
}%

\usepackage[unicode=true,colorlinks=true,citecolor=blue,urlcolor=blue]{hyperref}

\usepackage{bm}
\usepackage{color}
\usepackage{epsfig}

\usepackage[normalem]{ulem}

\renewcommand {\phi}{{\varphi}}
\newcommand {\rmi}{{\rm i}}

\newcommand {\e}{{\rm e}}

\graphicspath{{./figs/}{./}}

\begin{document}
\title{
Correlated relaxation and emerging entanglement in arrays of $\Lambda$-type atoms
}

\author{Denis Ilin}
\affiliation{School of Mathematical and Physical Sciences, University of Technology Sydney, Ultimo, NSW 2007, Australia}\affiliation{ Sydney Quantum Academy, Sydney, NSW 2000, Australia}

\author{Alexander V. Poshakinskiy}
\affiliation{ICFO-Institut de Ciencies Fotoniques, The Barcelona Institute of Science and Technology, 08860 Castelldefels (Barcelona), Spain}

\author{Alexander S. Solntsev}
\affiliation{School of Mathematical and Physical Sciences, University of Technology Sydney, Ultimo, NSW 2007, Australia}

\begin{abstract}
We present a theoretical framework for investigating the non-classical collective relaxation in arrays of $\Lambda$-type three-level atoms, where the two optical transitions coupled to a pair of orthogonal chiral or achiral waveguide modes. We demonstrate that the atomic entanglement emerges in the course of relaxation and persists in the final steady state of the system. We also reveal the entanglement of among the emitted photons as well as between the atoms and the photons. The presence and the degree of the entanglement depends crucially on the interatomic distance and the chirality of the waveguide. Our findings open a new way to engineer dissipation-induced entanglement.
\end{abstract}

\maketitle 
\date{\today}

\twocolumngrid

\section{Introduction}

\begin{figure}[t]
\centering
\includegraphics[width=0.8\columnwidth]{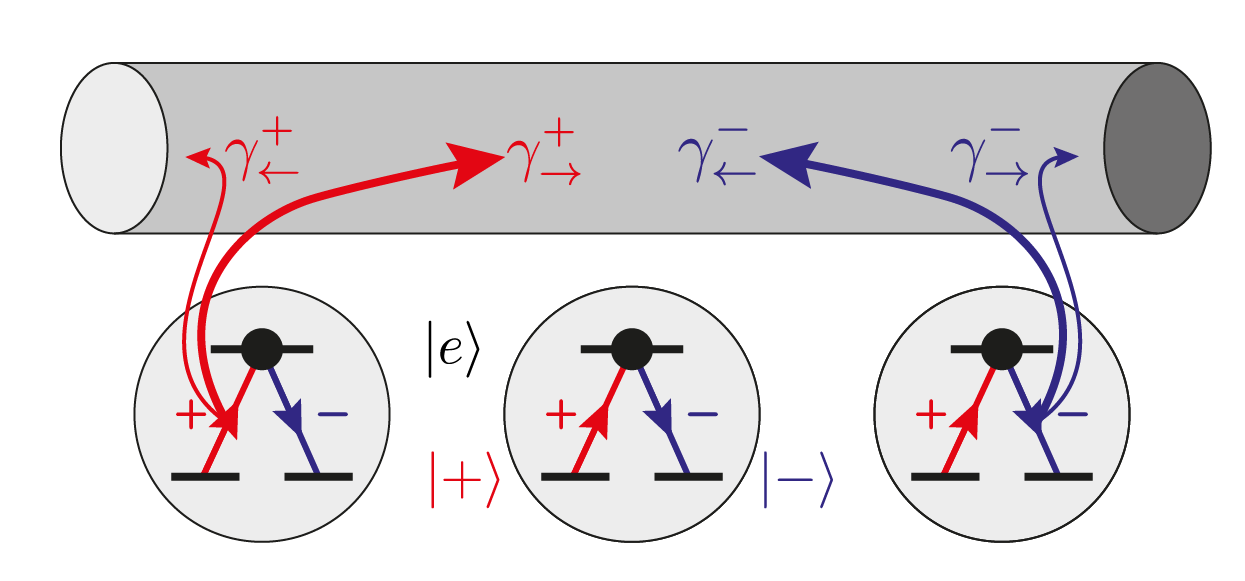}
\caption{An array of three-level $\Lambda$-type atoms with degenerate ground states $|\pm \rangle$,  coupled to a chiral waveguide.}\label{fig:system}
\end{figure}

The engineering of the interaction of light with quantum emitters has become a cornerstone of quantum optics and quantum information science~\cite{Lodahl2017, Solntsev2021}.
Of particular interest are waveguide quantum electrodynamical (WQED) setups, where quantum emitters like atoms or artificial atoms are coupled to a one-dimensional photonic mode~\cite{Roy2017,Chang2018,Sheremet2023}. This enables strong interactions between photons and atoms that are otherwise weak in free space. The platform provides a fertile ground for exploring fundamental quantum phenomena, such as non-classical photon correlations~\cite{Fan2007}, enabling quantum photon gates~\cite{Ralph2015}.
If several atoms are coupled to the same waveguide, their dynamics is described by collective modes which 
can have lifetimes that differ significantly from that of an individual atom~\cite{Sheremet2023} and complex spatial distributions~\cite{Poshakinskiy2021c,Poshakinskiy2021h}. Typically, the initial fast relaxation is governed by superradiant Dicke modes~\cite{Dicke1954,Yudson1984}, while  the subradiant fermionized  modes~\cite{Molmer2019,Ke2019,poshakinskiy2021} get populated at large times~\cite{Henriet2019}.    

Lately, the interest has shifted from conventional WQED systems with two-level atoms to those with multilevel emitters of various types~\cite{Witthaut2010,Das2018,Sun2023,Gao2024}. 
Among them, $\Lambda$-type three-level atoms have garnered particular interest due to their ability to store and manipulate quantum information~\cite{Li2018, Cai2023}. In a typical $\Lambda$-type system, the atom has two ground states and one excited state, with transitions between the ground states and the excited state mediated by photons with orthogonal polarizations. When $\Lambda$-type atom is in its ground state, it acts as a qubit that can store information for long times. Incoming photons can be used to manipulate the qubit state via single-photon Raman interaction~\cite{Pinotsi2008,Rosenblum2015}.
Previous studies have primarily focused on the behavior of a single $\Lambda$-type atom coupled to a waveguide, exploring phenomena such as the scattering of a single photon~\cite{Pinotsi2008,Zhong2023,Poddubny2024} or several photons~\cite{Ilin2024}, the realization of atom-photon quantum gates~\cite{Koshino2010,Li2012,Rosenblum2017,Bechler2018,Chan2023},
photon-mediated state transfer~\cite{Li2018}, and the generation of entangled photon states~\cite{Aqua2019,Gao2024,Chien2024}. 
However, the effects of collective dynamics in an array $\Lambda$-type atoms coupled to the same waveguide remain largely unexplored.

In this work, we develop a comprehensive theoretical framework to describe the collective relaxation dynamics for arrays of $\Lambda$-type atoms coupled to chiral or achiral waveguides. We focus on the properties of the final atomic state, where all atoms are in their ground states. We show that starting from uncorrelated atomic state, i.e., a product state, the collective relaxation can bring atoms to a final stationary state that is entangled. We also analyze the quantum correlations between the final atomic state and the state of the  emitted photons. Moreover, we show that the polarizations and propagation directions of the emitted photons are pairwise entangled.
Our findings reveal the intricate ways in which quantum correlations evolve in multi-atom systems and provide new insights into the control and manipulation of quantum states for practical applications. Our results not only offer a deeper understanding of entanglement dynamics but also highlight new pathways for designing quantum protocols that exploit the interplay between multi-level atomic structure, collective atom-waveguide interactions, and waveguide chirality.

\section{Model} 
We consider an array of $N$ identical $\Lambda$-type three-level atoms coupled to a two-mode waveguide, see Fig.~\ref{fig:system}. The  the Hamiltonian of the system reads
\begin{equation}
    \begin{aligned}
        H &=\sum\limits_{k}\sum_{\sigma = \pm}\omega_k\hat{a}^{\dagger}_{k,\sigma}\hat{a}_{k,\sigma} +\sum\limits_{j=1}^{N}\omega_0|e\rangle_j\langle e|_j 
        \\&+\sum\limits_{j=1}^{N}\sum\limits_{k}\sum\limits_{\sigma=\pm} g^{(\sigma)}_{k}  |e\rangle_j\langle \sigma|_j 
        \hat{a}_{k,\sigma}e^{\rmi kz_j} + \rm{H.c.},
    \end{aligned}
\end{equation}
where $\hat{a}_{k,\pm}$ are the bosonic operators for the photons with momentum $k$ for polarization $\pm$, which are assumed to have the same dispersion $\omega_k=c|k|$, 
$|\pm\rangle_j$ are the two degenerate ground states of $j$-th atom, $|e\rangle_j$ is its excited state, and $z_j$ is the atom coordinate along the waveguide. The transition between the excited state $|e \rangle$ and the ground states $|\pm \rangle$ has frequency $\omega_0$ and is coupled to waveguide photons of $\pm$ polarization with amplitude $g^{(\sigma)}_{k}$.

The peculiar feature of $\Lambda$-atoms is the $2^N$-fold degeneracy of their ground state. This allows for the qubits residing in a ground state to be entangled. Starting from a certain excited state, the array of $\Lambda$-atoms shall eventually relax to one of the ground states, or their mixture. The goal of this paper is to determine the final state and describe its properties, in particular, the emerging qubit entanglement.

To describe the relaxation of the system, we use the standard procedure and trace out the photonic degrees of freedom, assuming the Born-Makrov approximation. We assume no external pump, hence there is no explicit time dependence, and the dynamics is considered in the reference frame rotating with frequency $\omega_0$. The density matrix of qubits is governed by the dynamical Lindblad equation
\begin{align}
    \frac{d\rho}{dt}= \mathcal{L}[\rho] \equiv \sum_{ij,\sigma} \gamma_{ij}^{(\sigma)} (b_{j,\sigma} \rho b_{i,\sigma}^\dag  -b_{i,\sigma}^\dag b_{j,\sigma} \rho  )   + {\rm H.c.},
\end{align}
where we introduced the jump operators ${b}_{i,\sigma}=|\sigma\rangle_i\langle e|_i$ corresponding to emission of the $\sigma$-polarized photon, 
\begin{align}
    \gamma_{ij}^{(\sigma)} = \e^{\rmi k_0 |z_i-z_j|}\begin{cases}
    \gamma_{\rightarrow}^{(\sigma)}, & i>j \\
    \gamma_{\leftarrow}^{(\sigma)}, & i<j \\
    \frac12[\gamma_{\rightarrow}^{(\sigma)}+\gamma_{\leftarrow}^{(\sigma)}], & i=j
    \end{cases},
\end{align}
and $\gamma_{\rightleftarrows}^{(\sigma)} = (g_{\pm k_0}^{(\sigma)})^2/2c$ which describe the the decay rate of a single excited $\Lambda$-atom into the state $|\sigma \rangle$ during the emission of a photon with $k_0 = \omega_0/c$ in the direction $\rightleftarrows$.
We will focus on the case where the time-reversal invariance holds, so $\gamma_{\rightarrow}^{(+)} = \gamma_{\leftarrow}^{(-)} =\gamma$ and $\gamma_{\rightarrow}^{(-)} = \gamma_{\leftarrow}^{(+)} = s \gamma$, where the parameter $s$ quantifies the chirality of the atom-waveguide interaction. In principle, the chirality of the waveguide also leads to a dependence of the photon wave vector $k_0$ on the propagation direction and polarization. However, the corresponding optical phases can be removed by a local unitary transform of the ground atomic states, and thus do not affect the entanglement. 

Note that in the achiral case $s=1$, the system has a global SU(2) symmetry defined by the global transformation of the atomic ground states $|\sigma\rangle \to  \sum_{\sigma'}U_{\sigma\sigma'}|\sigma\rangle$, where $U$ is a unitary matrix.

 The evolution of the qubit density matrix is given by the map $\mathcal M_t[\rho_0] = \e^{\mathcal{L}t}[\rho_0]$, where $\rho_0$ is the initial state. In particular, the final state reads $\rho_\infty = \mathcal M_\infty[\rho_0]$. To evaluate $\mathcal M_\infty$, it is convenient to perform the eigenvalue decomposition  of $\mathcal L$, 
$ \mathcal{L} [\rho] = \sum_\lambda  \lambda U_\lambda {\rm Tr}( V^\dag_\lambda \rho) $,
where $U_{\lambda}$ and $V_\lambda$ are the eigenvectors of $\mathcal{L}$ and $\mathcal{L}^\dag$, respectively, that satisfy ${\rm Tr}( V^\dag_\lambda U_{\lambda'}) = \delta_{\lambda\lambda'}$. The final state is then defined by the projection onto the null space, $\mathcal M_\infty[\rho] = \sum_{\lambda=0}  U_\lambda {\rm Tr}( V^\dag_\lambda \rho)$~\cite{Albert2014}. 
The final density matrix can have nonzero elements only between the ground states. Therefore, the final state can be regarded as an $N$-qubit state.

\begin{figure}[t]
\centering
\includegraphics[width=0.48\textwidth]{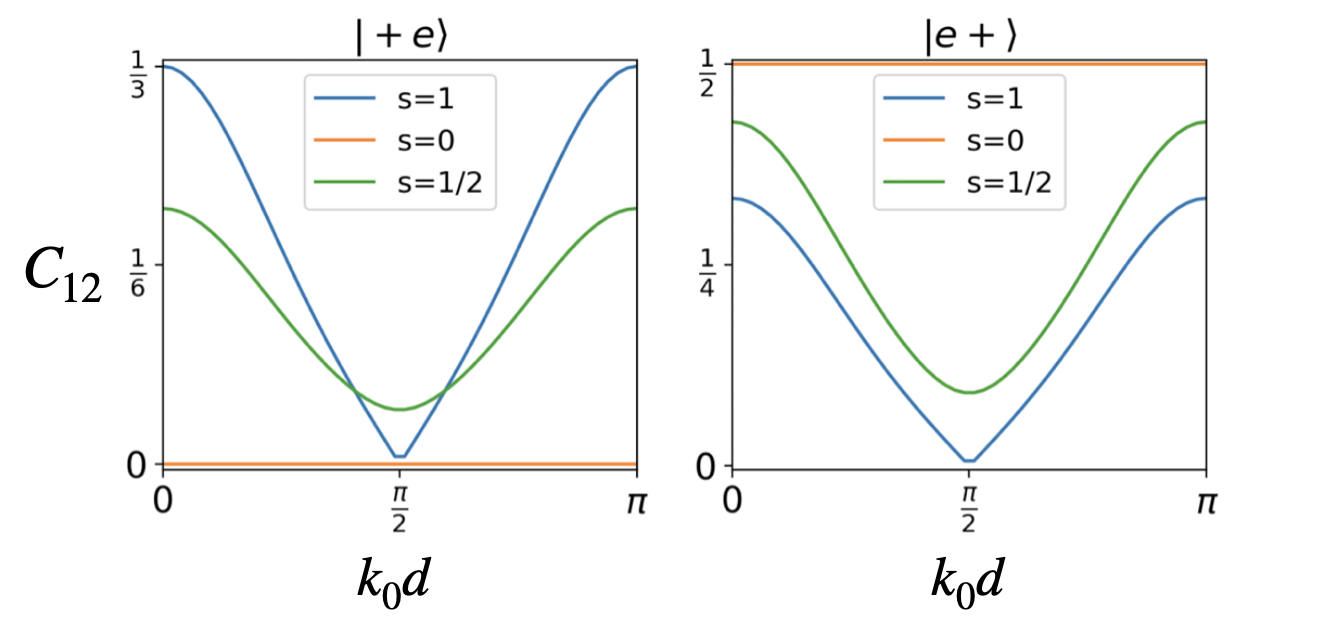}
\caption{Concurrence $C_{12}$ of the final state of $N=2$ atoms as a function of the spacing $d$ plotted for initial configurations $|+e\rangle$ and $|e+\rangle$. Parameter $s$ quantifies the chirality of the system, with $s=0$ being the ideally chiral case and $s=1$ the achiral case.}\label{fig:N2}
\end{figure}

\section{Correlated relaxation}\label{sec:atat}

When a single excited $\Lambda$-atom emits a photon, the final state reads
\begin{align}\label{eq:psi1}
\psi_\infty= \frac{a_{+}^\dag|+\rangle + a_{-}^\dag|-\rangle}{\sqrt2} ,
\end{align}
where the photonic operators for the emitted modes are
\begin{align}
a_{\pm} = \sum_{k} \frac{\sqrt{c\gamma}}{\omega_k-\omega_0+\rmi\gamma} \frac{a_{\pm k,\pm}+\sqrt{s}a_{\mp k,\pm}}{1+s} .
\end{align}
Note that the state Eq.~\eqref{eq:psi1} is the Bell state, and when the photon is traced out the final atom state is unpolarized, 
 $\rho_{\infty} =\tfrac12 \sum_{\sigma=\pm}|\sigma\rangle\langle \sigma| $. Now, we consider arrays of several atoms and show that the entanglement between the atoms can emerge even when the photons are traced out.


\subsection{Emerging entanglement of $N=2$ atoms}

\begin{figure}[t]
\centering
\includegraphics[width=0.48\textwidth]{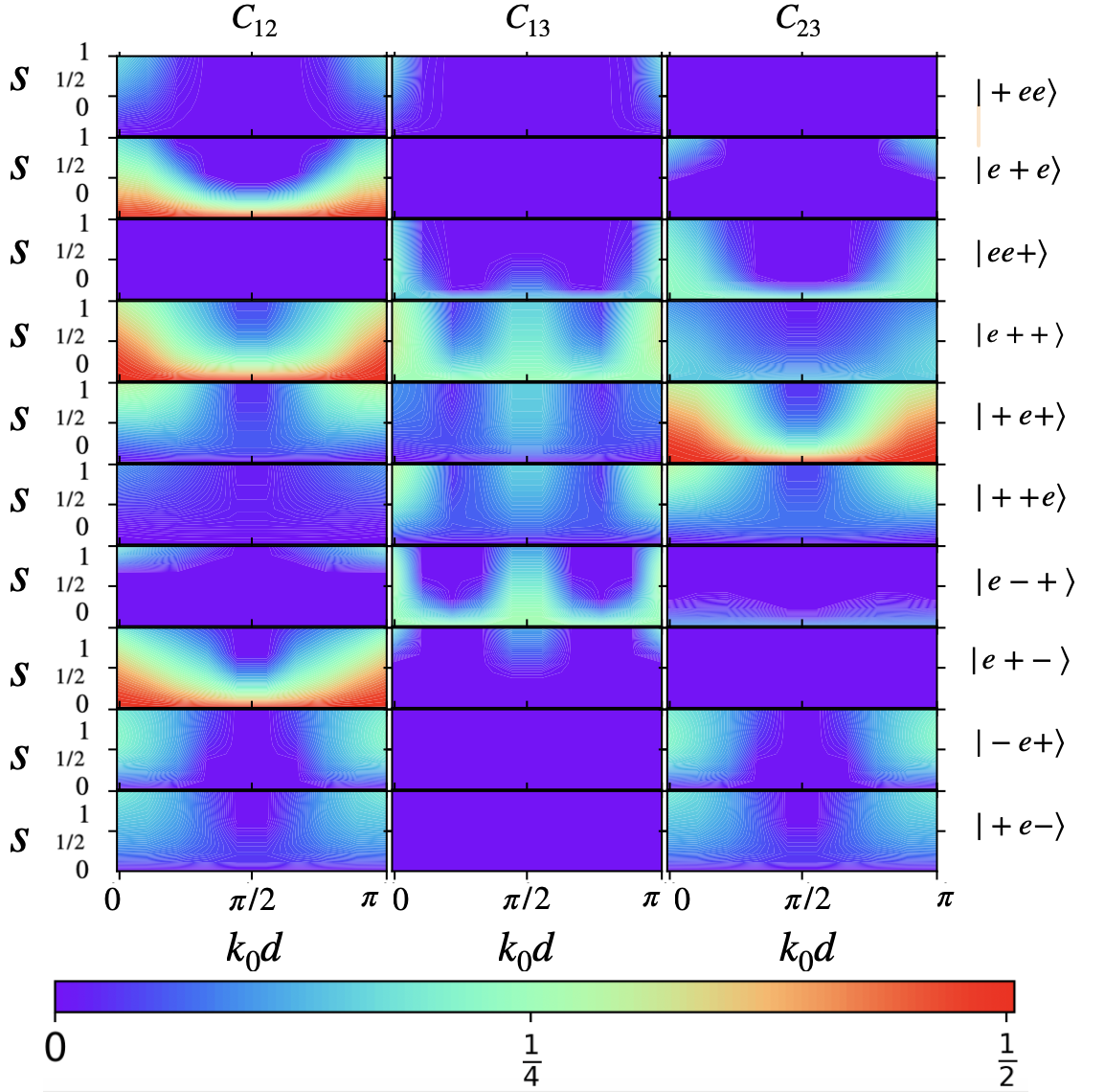}
\caption{Pairwise concurrences $C_{12}$, $C_{13}$ and $C_{23}$ for the final state of $N=3$ equidistant atoms, as a function of the inter-atomic distance $d$  and the chirality parameter $s$. Calculation is performed for all different non-trivial  initial configurations, indicated in the graph.}\label{fig:N3}
\end{figure}

\begin{figure}[t]
\centering
\includegraphics[width=0.48\textwidth]{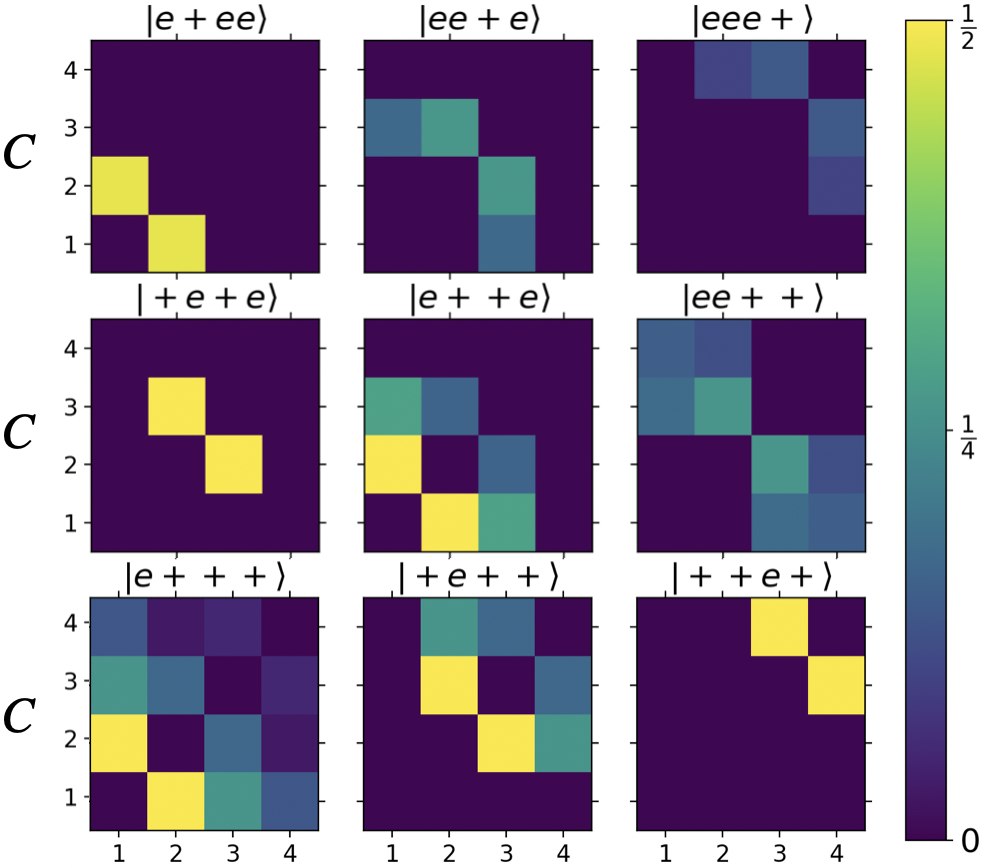}
\caption{The matrix of pairwise concurrences $C_{ij}$, plotted for different initial states of the array of $N=4$ atoms. Calculation is done for the ideally chiral case, $s=0$, and does not depend on the inter-atomic distances.
}\label{fig:N4_ch}
\end{figure}

If both atoms are initially excited, $\rho_0 = |ee\rangle\langle ee|$, the final state appears to be always separable. In particular, for the achiral case $s=1$, the global SU(2) symmetry imposes that the final state $\rho_\infty$ should give the same probability for all the states within the subspace with a certain total (ground state) spin $S$. The calculation yields
\begin{align}\label{eq:rhoee}
    \rho_\infty = \frac{P_1+ P_0 \sin^2 k_0d }{3+\sin^2 k_0d} \,,
\end{align}
where $P_0 = \tfrac12 [|+-\rangle -|-+\rangle][\langle +- |-  \langle -+ |]$
and $P_1 =1- P_0$ are the projectors onto the subspaces with $S=0$ and $1$, respectively. Even though $P_0$ describes an entangled state, the admixture of $P_1$ leads to the separability of the final state for any value of the inter-qubit distance $d$. Similarly, in the chiral case, $s=0$, the final state
\begin{align}\label{eq:rhoCee}
    \rho_\infty = \tfrac5{16} \big[&|++\rangle \langle ++|+ |--\rangle \langle --| \big] \nonumber\\
    + \tfrac18 &|-+\rangle \langle -+| + \tfrac14|B\rangle \langle B| \,,
\end{align}
where $|B\rangle = [|-+\rangle+\e^{2\rmi k_0d}|+-\rangle]/\sqrt2$, is also separable.

Next, we chose the initial state where only on of the atoms is excited.
In Fig.~\ref{fig:N2} we show the two-qubit concurrence $C_{12}$ calculated  as a function of the inter-qubit distance for different chirality parameters $s$ and the initial states $|e+\rangle$ or $|+e\rangle$. Note that the initial states $|e-\rangle$ or $|-e\rangle$ yield the same values of $C_{12}$ as $|+e\rangle$ or $|e+\rangle$, respectively.

We start the analysis from the ideally chiral case $s=0$  (orange curves), as this allows one to easily track the possible paths of the emitted photon. For the $|+e\rangle$ initial state, the relaxation of the second atom cannot affect the first one. Indeed, if the second atom emits a $\sigma=+$ photon, it propagates to the right and does not encounter the first atom, and if a $\sigma=-$ photon is emitted, it passes through the first atom in the $|+\rangle$ without interaction. As a result, the final state is a product state. 
If the initial state is $|e+\rangle$, the first atom can emit a $\sigma=+$ photon in the right direction, which is then scattered by the second atom, or a $\sigma=-$ photon in the left direction. The final state of the atoms and the photon reads
\begin{align}
    \psi_\infty &= \sum_{k} \frac{\sqrt{c\gamma}}{\omega_k-\omega_0+\rmi\gamma}\Big\{ a_{-k,-}^\dag|-+\rangle \\ \nonumber
    &+ \e^{2\rmi k_0d} \Big[t(\omega) a_{k,+}^\dag|++\rangle + r(\omega) a_{-k,-}^\dag|+-\rangle \Big]\Big\} \,.
\end{align}
where $r(\omega) = t(\omega)-1 = -\rmi\gamma/(\omega-\omega_0+\rmi\gamma)$ are the single photon reflection and transmission coefficients for a chiral $\Lambda$-atom~\cite{Pinotsi2008,Ilin2024}.
After averaging over the photons, we find the density matrix of the atoms
\begin{align}
    \rho_\infty = \tfrac14 |++\rangle \langle ++| + \tfrac14|-+\rangle \langle -+| + \tfrac12|B'\rangle \langle B'| \,,
\end{align}
where $|B'\rangle = [|-+\rangle-\e^{2\rmi k_0d}|+-\rangle]/\sqrt2$ is the entangled Bell state. As a result, the concurrence of the final atomic state is equal to $1/2$.

In the achiral case $s=1$ (blue curves), the concurrence of the final state $C_{12}$ is the same for the two initial conditions, due to the parity and the global SU(2) symmetry. Instead, $C_{12}$ has as strong dependence of the inter-qubit distance $d$.
The concurrence reaches the maximal value 1/3 near the Bragg condition, $k_0d=0,\pi$, and vanishes at the anti-Bragg condition, $k_0 d=\pi/2$. 

\subsection{Emerging entanglement of  $N>2$ atoms}

Now we consider the entanglement that emerges during the relaxation of larger atomic arrays with $N=3,4$. Although the characterization of the entanglement of $N>2$ atoms is not straightforward, its main features remain the same as in the considered above case of $N=2$ atoms. In paticular, if in the initial state all atoms were excited, the atoms in the final state are not entangled pairwise, and we have also checked that all bipartitions have positive partial transpose (PPT). 
Therefore, we focus on the initial states where one or several atoms are in ground state.  

In Fig.~\ref{fig:N3} we plot pairwise concurrences $C_{12}$, $C_{13}$ and $C_{23}$ of the final state of
of $N=3$ equidistant atoms for all nontrivial initial states. The concurrence reaches maximal values $C=1/2$ for certain configurations in the chiral case $s=0$, when it does not depend on the inter-atomic distance. In the achiral case, $s=1$, the dependence on $d$ is significant. The concurrences $C_{12}$ and $C_{23}$ are maximal for the Bragg spacing, $d=0,\pi$, while $C_{13}$ is stronger in the anti-Bragg case, $d=\pi/2$.


In Fig.~\ref{fig:N4_ch} we plot the pairwise entanglement diagrams for $N=4$ atoms in the ideal chiral case $s=0$. Here, we focus our attention on the initial states where all atoms are excited or in the $|+\rangle$ ground state.  It is easy to identify the general feature of these diagrams: The entanglement originates from the pairs of atoms that had the initial state of the form $|\ldots e \ldots +\ldots\rangle$, and is suppressed with increase of the distance between the atoms. The most demonstrative is the initial state $|e+++\rangle$. When a right-propagating $+$ photon is emitted by the first atom, it can be scattered by one of the atoms in the $|+\rangle$ state, and the probability to reach a particular atom decreases with the increase of the atom index. As a result, the photon gets entangled with all the atoms, forming a generalized $W$-class state~\cite{Ilin2024}. When the photon is traced out, the all-to-all entanglement of the atoms persists, with the strongest entanglement between the first two atoms, $C_{12}=1/2$.  Several other initial conditions, for example, $|+e++\rangle$, $|++e+\rangle$, $|e++e\rangle$, $|+e+e\rangle$, $|e+ee\rangle$ can be regarded as a truncated version of the described scheme.
Another interesting initial condition is $|eee+\rangle$. Then, each of the first three atoms gets entangled to the fourth atom via a different photon, which suppresses the overall atom entanglement.

\section{Emission upon relaxation}

When a single $\Lambda$-atom emits a photon, the photon polarization and the final atomic state are entangled and described by the pure Bell state Eq.~\eqref{eq:psi1}. When not all of the emitted photons are considered, the joint state of the atoms and detected photons is a mixed state, but it can still feature some entanglement. In particular, now we suppose that a single photon is detected and consider the correlations it has with the atomic state. To this end, we calculate the final joint atoms-photon density matrix. If a single photon detected at time $t$, the density matrix reads $
\rho_{\sigma,\delta;\sigma',\delta'}(t) = \mathcal{M}_\infty[ b_{\sigma,\delta} \mathcal{M}_t(\rho_0) b_{\sigma',\delta'}^\dag ]$,
which yields the detection-time-averaged density matrix
\begin{align}\label{eq:rho1}
\rho_{\sigma,\delta;\sigma',\delta'} = -\frac1{N_0}\,\mathcal{M}_\infty[ b_{\sigma,\delta} \mathcal{L}^{-1}(\rho_0) b_{\sigma',\delta'}^\dag ]
\end{align}
where $\sigma, \sigma' = \pm$ is the polarization of the detected photon, $\delta,\delta' = \rightleftarrows$ is its propagation direction, 
\begin{align}
b_{\sigma,\rightleftarrows} = \sqrt{\gamma_{\rightleftarrows}^{(\sigma)}} \sum_j b_j^{(\sigma)} \e^{\mp\rmi k_0 z_j} \,.
\end{align}
and $N_0$ is the total number of emitted photons, which is equal to the number of excited atoms in the initial state.
Note that the divergence of the map $\mathcal{L}^{-1}$ in the ground subspace is eliminated by the subsequent application of the lowering operator $b_{\sigma,\delta}$ in Eq.~\eqref{eq:rho1}.

\begin{figure}[t]
\centering
\includegraphics[width=0.48\textwidth]{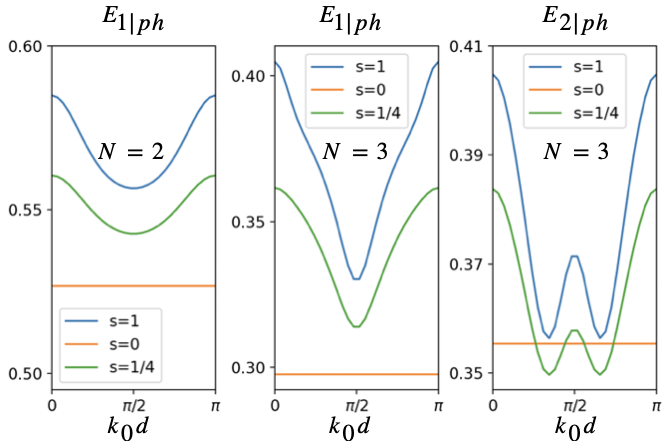}
\caption{Logarithmic negativity $E_{{\rm ph}|j}$ of the reduced final state of the $j$-th atom and a detected photon. Calculation is performed for  $N=2$, 3 atoms that were initially in the $|e
\rangle^N$ state for different values of the chirality parameter $s$, indicated in the plot.}\label{fig:Eap}
\end{figure}

\begin{figure}[t]
\centering
\includegraphics[width=0.48\textwidth]{
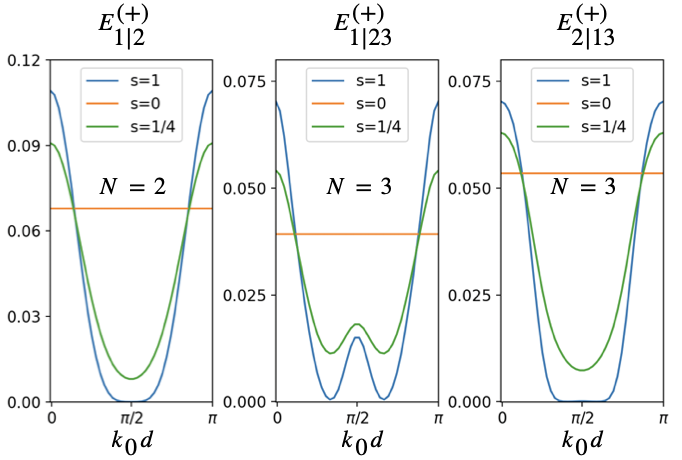}
\caption{Logarithmic negativity of the final state of $N=2$, 3 atoms that were initially in the $|e
\rangle^N$ state, conditioned by the detection of a $+$ photon. In the case $N=3$, two non-equivalent  bipartitions $1|23$ and $2|13$ of the atomic state are considered. Calculation is performed for different values of the chirality parameter $s$, indicated in the plot.}\label{fig:Eatc}
\end{figure}

\subsection{Atom-photon entanglement}

First, we study the entanglement between the detected photon and each of the qubits. Since the photonic state resides in a 4-dimensional space (describing two polarizations and two propagation directions), this entanglement cannot be described by conventional qubit measures. Instead, we use the logarithmic negativity $E_{A|B}= \log_2 {\rm Tr\,}| \rho_{AB}^{T_A}|$, where ${T_A}$ denotes the partial transpose with respect to subsystem $A$, as an entanglement measure between subsystems $A$ and $B$---specifically, an atom and an photon. Nonzero logarithmic negativity is a sufficient condition for the entanglement and establishes the upper bound of the distillable entanglement~\cite{Vidal2002}. Figure~\ref{fig:Eap} shows the logarithmic negativity $E_{{\rm ph}|j}$ of the reduced state of the detected photon and the $j$-th atom, with all other atoms traced out. One can see that the entanglement persists for the whole range of parameters and depends rather weakly on chirality $s$, the array period $d$, and the selected atom. Clearly, entanglement reduces when the number of atoms increases from $N=2$ to $N=3$, which might be interpreted as a result of entanglement monogamy~\cite{Coffman2000}. 

\subsection{Post-selected atom-atom entanglement upon photon detection}

Because the detected photon and the atomic array are entangled, measurement of the photon state can induce entanglement within the atoms. Suppose that we have detected a photon of a certain polarization $\sigma$. Then, the state of the atoms $\rho^{(\sigma)}$ is described by  $\sum_\delta\rho_{\sigma,\delta;\sigma,\delta}$, which should be renormalized. Figure~\ref{fig:Eatc} shows the logarithmic negativity of the final state of $N=2$, 3 atoms that were initially in the $|e\rangle^N$ state, conditioned by the detection of a $+$ photon. 
The conditioned final atomic states can also be obtained from the unconditioned states, described in Sec.~\ref{sec:atat}, by weighting the matrix elements with the probabilities of the $+$ photon detection.
In particular for $N=2$, $|--\rangle\langle--|$ matrix element should be eliminated, since that state corresponds to the emission of two $-$ polarized photons,  the states
$|\pm \mp \rangle\langle \pm \mp|$, $|\pm \mp \rangle\langle \mp\pm|$ should be weighted with the probability $1/2$, and the $|++\rangle\langle ++|$ state with probability 1. For the chiral case, $s=0$, using Eq.~\eqref{eq:rhoee}, we get the conditioned final state 
\begin{align}
    \rho^{(+)} = \tfrac58 |++\rangle \langle ++| + \tfrac18|-+\rangle \langle -+| + \tfrac14|B'\rangle \langle B'| 
\end{align}
that is characterized by the concurrence $C_{12}^{(+)} = 1/4$ and logarithmic negativity $E_{1|2}^{(+)} \approx 0.068$. In the achiral case, $s=0$, we use \eqref{eq:rhoCee} and get the final state that is characterized by the concurrence 
\begin{align}
C_{12}^{(+)} = \frac{\cos^2 k_0 d}{4-\cos^2 k_0d} \,,
\end{align}
which reaches the maximal value $C_{12}^{(+)} =1/3$ ($E_{1|2}^{(+)} \approx 0.11$) in the Bragg case, $k_0d = 0, \pi$. We recall that without the measurement of the photon, the final states Eqs.~\eqref{eq:rhoee}-\eqref{eq:rhoCee} were not entangled; the measurement transfers the atom-photon entanglement into the atom-atom entanglement.

For $N=3$ atoms, the calculation shows that the atoms are not pairwise entangled, i.e., all pairwise concurrences $C_{12}$, $C_{13}$ and $C_{23}$ vanish. Nevertheless, the three atom state is entangled, which is revealed by the logarithmic negativities for the bipartitions, $E_{1|23}^{(+)} = E_{3|12}^{(+)}$ and $E_{2|13}^{(+)}$ shown in Fig.~\ref{fig:Eatc}.

\subsection{Photon-photon entanglement}

\begin{figure}[t]
\centering
\includegraphics[width=0.48\textwidth]{
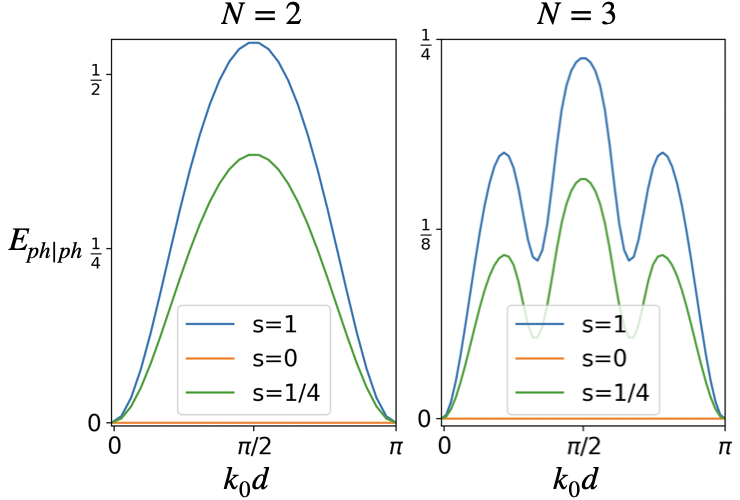}
\caption{Entanglement negativity of a pair of photons emitted by an array of $N=2$ or 3 atoms, that were initially in the state $|e\rangle^N$. Calculation is done for different values of the chirality parameter $s$ indicated in the plot. }\label{fig:2phot}
\end{figure}

Finally, we study correlations between the emitted photons. The polarizations of the two detected photons determine the final state of a pair of atoms. Since in the final state, which is reached starting from $|e\rangle^N$, the atoms are not entangled, as was shown previously, nor is the polarization state of the photons. However, if the direction of the photon emission is taken into account, the pairwise photon entanglement arises. To describe it, we introduce the two-photon density matrix
\begin{align}\label{eq:rho2ph}
&\rho_{\sigma_1,\delta_1,\sigma_2,\delta_2;\sigma_1',\delta_1',\sigma_2',\delta_2'} \\ \nonumber
&\ \ = \frac2{N_0(N_0-1)}\, {\rm Tr}\{ b_{\sigma_2,\delta_2}\mathcal{L}^{-1}[ b_{\sigma_1,\delta_1} \mathcal{L}^{-1}(\rho_0) b_{\sigma_1',\delta_1'}^\dag ] b_{\sigma_2',\delta_2'}^\dag \} \,
\end{align}
which is averaged over the detection times of the two photons and the final state of the atoms. 

To characterize pairwise photon entanglement, we calculate the logarithmic negativity $E_{\rm ph|ph}$ of the two-photon state Eq.~\eqref{eq:rho2ph}. Figure~\ref{fig:2phot} shows the result for the arrays of $N=2$ and $N=3$ atoms. For the chiral case, $s=0$ (orange curves), the entanglement negativity vanishes, because the photon emission directions are locked to their polarizations, and the latter are not entangled, as discussed above. In achiral system (blue curves) the negativity $E_{\rm ph|ph}$ is finite, proving the presence of the pairwise photon entanglement, except for the Bragg-spaced arrays with $k_0d=0,\pi$. The pairwise photon entanglement negativity reduces when the number of atoms is increased form 2 to 3, as a result of entanglement monogamy and the entanglement with the third emitted photon which is not measured.

\section{Conclusions}

In this study, we have considered theoretically the relaxation of the arrays $\Lambda$-type atoms. We find that starting form a separable excited state, the relaxation leads the system to a final steady state of atoms and emitted photons, where 
the atom-atom, atom-photon, and photon-photon entanglement emerges. The entanglement has a rather complex dependence on the spacial arrangement of the atoms and to the chirality of the waveguide. In particular, the emerging atom-atom entanglement is typically stronger in the cases of  chiral waveguide or Bragg-spaced atomic array. In contrast, photon-photon entanglement vanishes in these cases, while it is maximal for anti-Bragg-spaced array and an achiral waveguide.

Our results open new methods for entanglement generation through collective dissipation, which may find their application in quantum technologies, and provide avenues for future research. While our consideration was limited to rather small systems with up to $N \leq 4$ $\Lambda$-type atoms, which allow for the exact calculation of the evolution and simple analysis of the entanglement in the final state, the study of entanglement propagation and scaling in  arrays of many atoms would be of considerable interest in the future, despite being numerically hard. Another generalization might be by incorporating multiple coupled waveguides ~\cite{Solntsev2016, Jiang2022}, which would further expand the landscape of quantum phenomena.

\bibliography{lambda.bib}

\end{document}